\documentclass[aps,preprint,showpacs]{revtex4-1}
\usepackage{amssymb}
\usepackage{mathrsfs}
\usepackage{lipsum}
\usepackage{graphicx}
\usepackage[caption=false]{subfig}
\usepackage{mathtools}
\usepackage{epstopdf}
\usepackage{xcolor}
\usepackage[colorlinks, linkcolor=red, anchorcolor=green, citecolor=blue, urlcolor=violet]{hyperref}
\DeclareGraphicsExtensions{.pdf,.jpg,.png,.eps}
\usepackage[toc,page,title,titletoc,header]{appendix}
\begin{document}
	
	\title{Conclusive nonlinear phase sensitivity limit for a Mach-Zehnder interferometer with single-mode non-vacuum inputs}

	\author{Jian-Dong Zhang}
	\affiliation{School of Physics, Harbin Institute of Technology, Harbin 150001, China}
	\author{Zi-Jing Zhang}
	\email[]{zhangzijing@hit.edu.cn}
	\affiliation{School of Physics, Harbin Institute of Technology, Harbin 150001, China}
	\author{Jun-Yan Hu} 
	\affiliation{School of Physics, Harbin Institute of Technology, Harbin 150001, China}
	\author{Long-Zhu~Cen} 
	\affiliation{School of Physics, Harbin Institute of Technology, Harbin 150001, China}
	\author{Yi-Fei Sun} 
	\affiliation{School of Physics, Harbin Institute of Technology, Harbin 150001, China}
	\author{Chen-Fei Jin}
	\email[]{jinchenfei@hit.edu.cn}
	\affiliation{School of Physics, Harbin Institute of Technology, Harbin 150001, China}
	\author{Yuan Zhao}
	\email[]{zhaoyuan@hit.edu.cn}
	\affiliation{School of Physics, Harbin Institute of Technology, Harbin 150001, China}

	\date{\today}
	
	\begin{abstract}
		Many works have stated that nonlinear interactions can improve phase sensitivity beyond the Heisenberg limit scaling of $1/N$ with $N$ being the mean photon number. 
		This raises some open questions---among them the conclusive sensitivity limits with respect to single-mode inputs.
		Namely, when one of two inputs is vacuum, is there a shot-noise-style sensitivity bound on a nonlinear Mach-Zehnder interferometer?
		Within the reach of second-order nonlinear phase shifts, we make an attempt to provide an answer to this question.
		Based upon phase-averaging approach, this puzzle is partially resolved with careful calculations of the quantum Fisher information regarding three kinds of common inputs: Gaussian states, squeezed number states, and Schr\"odinger cat states.
		The results suggest that shot-noise-style sensitivity limit is no longer available, and the ideal candidate is squeezed vacuum.
		
	\end{abstract}

	\maketitle

	\section{Introduction}
	
	Optical interferometry is$\--$and always was$\--$an essential component of the field of precision measurements at the microscale. 
	In recent years, classical interferometry fails to keep pace with ever-growing performance requirements in resolution and sensitivity.
	Within this context, utilizing exotic quantum properties to improve performance is an effectual way.
	Related to this, quantum precision measurements and the corresponding theoretical science, quantum metrology, emerge as focal points.
	Owing to the importance of quantum metrology shown in a great deal of areas, it has become one of the most influential quantum technologies \cite{PhysRevLett.96.010401,Giovannetti2011Advances,PhysRevLett.109.233601,RevModPhys.89.035002}.

	For many years linear phase estimation has cast its spell over researchers, for numerous parameters of interest can be mapped on an unknown phase in an optical interferometer.
	Breaking the classical limits$\--$Rayleigh diffraction limit in resolution and shot-noise limit in sensitivity$\--$is the primary task.
	As another branch of research in the field of quantum metrology, nonlinear phase estimation \cite{PhysRevLett.105.010403,PhysRevLett.101.040403,PhysRevA.66.013804,PhysRevA.90.063838} has also gained a lot of attention.
	Unlike well-established linear phase estimation, the studies on nonlinear phase estimation face some conceptual and mathematical difficulties.
	As a consequence, there are some open questions on phase sensitivity to be discussed.
	The two topical discussions among them are the shot-noise limit and Heisenberg limit, which are applicable for nonlinear phase estimation.

	Of the linear phase estimation, an interferometer driven by arbitrary single-mode inputs gives the same sensitivity, which is referred to as the shot-noise limit \cite{PhysRevD.23.1693}.
	Regarding nonlinear phase estimation, an inspired question on \emph{whether there exists a shot-noise-style sensitivity limit for single-mode inputs} naturally arises.
	Physically, this limit is related to the optimal strategy after traversing all possible positive operate valued measures (POVMs).
	Mathematically, it is given by a proper quantum Fisher information (QFI) \cite{PhysRevLett.72.3439} calculation, $\Delta \varphi = 1/\sqrt{\cal F_{\rm Q}}$.
	Jarzyna and Demkowicz-Dobrza\'nski pointed out that an ill-considered direct calculation to the QFI may give rise to a sensitivity limit with overestimation \cite{PhysRevA.85.011801}.
	More recently, with regard to single-mode inputs, Takeoka \emph{et al.} and You \emph{et al.} demonstrated that this overestimation can be eliminated in SU(2) and SU(1,1) interferometers, respectively, through the use of phase-averaging approach \cite{PhysRevA.96.052118,PhysRevA.99.042122}.
	For this reason, in this paper we scale this approach to nonlinear phase estimation and address the sensitivity limit for single-mode inputs.

	The remaining sections of this paper are organized as follows. 
	Section \ref{s2} introduces our estimation protocol and the fundamental principle of phase-averaging approach.
	In Sec. \ref{s3}, we take advantage of phase-averaging approach to calculate the QFIs of common single-mode inputs, and the results are analyzed and discussed.
	Finally, we summarize our work with a brief conclusion in Sec. \ref{s4}.

	\section{Estimation protocol and phase-averaging approach}
	\label{s2}
\begin{figure}[htbp]
		\centering\includegraphics[width=8cm]{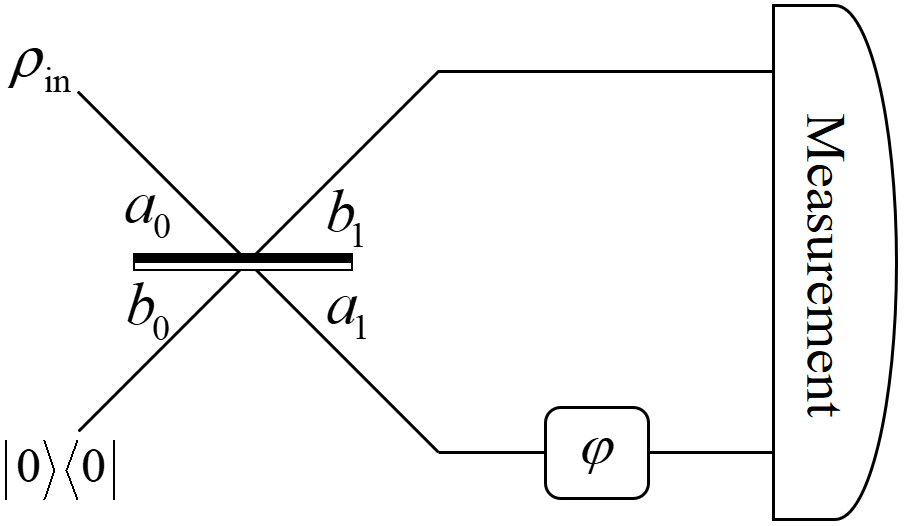}
		\caption{Schematic diagram of the estimation protocol for a second-order nonlinear phase $\varphi$.}
		\label{f1}
	\end{figure}

	Our protocol resembles a Mach-Zehnder interferometer but with the linear phase replaced by nonlinear phase, as illustrated in Fig. \ref{f1}.
	An arbitrary state and a vacuum one are injected into a 50/50 beam splitter from two different input ports.
	After the action of the beam splitter, all photons are redistributed to two modes, where $\hat a_0$ ($\hat b_0$) and $\hat a_1$ ($\hat b_1$) are the operators with respect to mode $A$ ($B$) before and after the beam splitter.
	The nonlinear phase shift $\varphi$ induced by a Kerr-type medium is imprinted on each photon passing through mode $A$.
	Regarding a second-order nonlinear phase shift, its operator form can be expressed as $\hat U_\varphi =\exp[ i (\hat a_1^\dag \hat a_1^{})^2\varphi ] $.
	Finally, one can perform a special POVM, from which the phase $\varphi$ can be estimated.

	Throughout this paper, the sensitivity limit for above estimation protocol is our interest.
	QFI-only calculations hold true for all POVMs, certainly with possibly hidden resources.
	These resources may weaken the tightness of QFI as conventional POVMs contain no hidden resources providing external power or phase reference.
	Under this circumstance, one is entitled to capitalize on the phase-averaging approach.
	In what follows, we direct our attention to the QFI calculation with this approach.
	
	According to the framework for the phase-averaging approach, the phase information $\theta$ in the input is initialized via operators ${\hat V_A} = \exp ( {i{{\hat a}^\dag }\hat a\theta } )$ and ${\hat V_B} = \exp ( {i{{\hat b}^\dag }\hat b\theta } )$,
	\begin{equation}
	\bar \rho  = \int_0^{2\pi } {\hat V_A^{}\hat V_B^{}{\rho _{\rm {in}}}\otimes{\left| 0 \right\rangle \left\langle 0 \right|} \hat V_A^\dag \hat V_B^\dag d\theta }, 
	\end{equation}
	where the input in mode $A$ can be expressed as:
	\begin{equation}
	{\rho _{{\rm{in}}}} = \sum\limits_{n,m}^\infty  {c_n^{}c_m^ * \left| n \right\rangle \left\langle m \right|}.
	\end{equation}
	Upon completing such treatment, the input turns out to be
	\begin{equation}
	\bar \rho  = \sum\limits_n^\infty  {{p_n}\left| n \right\rangle \left\langle n \right|}  \otimes \left| 0 \right\rangle \left\langle 0 \right|
	\label{e1}
	\end{equation}
	with ${p_n} = {\left| {{c_n}} \right|^2}$ being the probability of emerging $n$ photons in the input.
	Regarding the input in Eq. (\ref{e1}) and phase operator $\hat U_\varphi =\exp[ i (\hat a_1^\dag \hat a_1^{})^2\varphi ] $, the QFI can be calculated through some bosonic algebra (see Appendix for the details)
	\begin{equation}
	{{\cal F}_{\rm Q}}\left( {{\bar \rho}} \right) = \sum\limits_n^\infty {{p_n}} {{\cal F}_{\rm Q}}\left( {\left| {{n,0}} \right\rangle } \right)
	\label{QFI}
	\end{equation}
	with
	\begin{equation}
	{\cal{F}_{\rm Q}}\left( {\left| {n,0} \right\rangle } \right) = \sum\limits_n {4{p_n}} \left\{ {\langle n,0|{{\left[ {\hat U_{{\rm{BS}}}^\dag {{(\hat a_1^\dag \hat a_1^{})}^2}\hat U_{{\rm{BS}}}^{}} \right]}^2}\left| {n,0} \right\rangle  - {{\left| {\langle n,0|\hat U_{{\rm{BS}}}^\dag {{(\hat a_1^\dag \hat a_1^{})}^2}\hat U_{{\rm{BS}}}^{}\left| {n,0} \right\rangle } \right|}^2}} \right\}.
	\label{QFI_n}
	\end{equation}
	The operator relation between the output and input modes of the beam splitter can be described as:
	\begin{align}
	\left( {\begin{array}{*{20}{c}}
		{\hat a_1^\dag }  \\
		{\hat b_1^\dag }  \\
		\end{array}} \right) = \frac{1}{{\sqrt 2 }}\left( {\begin{array}{*{20}{c}}
		1 & { - i}  \\
		{ - i} & 1  \\
		\end{array}} \right)\left( {\begin{array}{*{20}{c}}
		{\hat a_0^\dag }  \\
		{\hat b_0^\dag }  \\
		\end{array}} \right).
	\end{align}
	Based on this transformation, the result of Eq. (\ref{QFI_n}) is found to be 
	\begin{equation}
	{\cal{F}_{\rm Q}}\left( {\left| {n,0} \right\rangle } \right) = \frac{n}{2}\left( {n + 1} \right)\left( {2n - 1} \right)
	\label{e7}
	\end{equation}
	Combining Eqs. (\ref{QFI}) and (\ref{e7}), the QFI of our protocol is recast as:
	\begin{eqnarray}
	{{\cal F}_{\rm Q}}\left( {{\bar \rho}} \right) =\frac{1}{2} \sum\limits_n^\infty {{p_n}} n\left( {n + 1} \right)\left( {2n - 1} \right).
	\label{e8}
	\end{eqnarray}
	
	It is by no means easy to obtain a mathematically tractable expression instead of an infinite series.
	As a result, we utilize a numerical method by terminating the series at a fidelity in excess of $0.99$. 
	The fidelity between two density matrices is defined as
	\begin{eqnarray}
    F\left( {\bar \rho ,\bar \rho '} \right): = {\left[ {{\mathop{\rm Tr}\nolimits} \left( {\sqrt {\sqrt {\bar \rho } \bar \rho '\sqrt {\bar \rho } )} } \right)} \right]^2}
	\end{eqnarray}
	with $\bar \rho '$ being the density matrix after truncated operation. 
	
	Of the linear phase estimation, the QFI in Eq. (\ref{e7}) is $n$; accordingly, the QFI in Eq. (\ref{e8}) is equal to the mean photon number $\cal N$.
	That is, the sensitivity limit is shot-noise limit, regardless of photon distribution in the input.
	However, there is no a shot-noise-style sensitivity limit in nonlinear phase estimation, since 
	QFI in Eq. (\ref{e8}) depends on the probability of photon number.
	
	\section{QFI calculations based on phase-averaging approach}
	\label{s3}
	At the end of the preceding section, the QFI based on phase-averaging approach is analyzed.
	Here we take advantage of this approach to discuss three kinds of common inputs: Gaussian states, squeezed number states, and Schr\"odinger cat states.
	\subsection{Gaussian states}
	Gaussian states, whose Wigner functions follow the Gaussian distribution, are crucial resources to the field of continuous-variable quantum information processing. 
	Related to this, many states play important roles in quantum metrology, such as single- and two-mode squeezed vacuum states \cite{PhysRevLett.111.173601,PhysRevLett.104.103602}.
	An arbitrary single-mode Gaussian state can always be written by a squeezed displaced thermal state \cite{RevModPhys.84.621}, defined as
	\begin{equation}
	{\rho _{\rm g}} = \hat S\left( r \right)\hat D\left( \alpha  \right){\rho _{\rm t}}{\hat D^\dag }\left( \alpha  \right){\hat S^\dag }\left( r \right),
	\end{equation}
	where  ${\rho _{\rm t}} = \sum\nolimits_n {{{{{ n}_{\rm th}^n}\left| n \right\rangle \left\langle n \right|} \mathord{\left/
				{\vphantom {{{{\bar n}^n}\left| n \right\rangle \left\langle n \right|} {{{\left( {n_{\rm th}^{} + 1} \right)}^{n_{\rm th}^{} + 1}}}}} \right.
				\kern-\nulldelimiterspace} {{{\left( {n_{\rm th}^{} + 1} \right)}^{n + 1}}}}} $ stands for the density matrix of a thermal state with $n_{\rm th}^{}$ photons on average, $\hat S\left( r \right) = \exp [ {{{( {{\xi ^ * }\hat a_0^2 - \xi \hat a_0^{\dag 2}} )}/2}}]$ and $\hat D\left( \alpha  \right) = \exp ( {\alpha \hat a_0^\dag  - {\alpha ^ * }\hat a_0^{}})$ denote squeezing operator and displacement operator, respectively, with $\alpha  = \left| \alpha  \right|{e^{i\phi }}$ and $\xi  = r{e^{i\chi }}$.
	The squeezing operator produces the following Bogoliubov transformation
	\begin{align}
	{{\hat S}^\dag }\left( r \right){{\hat a}_0^{}}\hat S\left( r \right) &= {{\hat a}_0^{}}\cosh r - e^{i\chi}\hat a_0^\dag \sinh r, \\
	{{\hat S}^\dag }\left( r \right)\hat a_0^\dag \hat S\left( r \right) &= \hat a_0^\dag \cosh r - e^{-i\chi}{{\hat a}_0^{}}\sinh r.
	\end{align}
	The action of the displacement operator over the creation and annihilation operators is given by
	\begin{align}
	{{\hat D}^\dag }\left( \alpha \right){{\hat a}_0}\hat D\left( \alpha \right) &= {{\hat a}_0} + \alpha, \\
	{{\hat D}^\dag }\left( \alpha \right)\hat a_0^\dag \hat D\left( \alpha \right) &= \hat a_0^\dag  + \alpha^{*}. 
	\end{align}

	To such a Gaussian state there corresponds the mean photon number,
	\begin{eqnarray}
	{{\cal N}_{\rm g}} =\frac{1}{2}\left[\left( {2{n_{{\rm{th}}}} + 1} \right)\cosh \left( {2r} \right) + 2{\left| \alpha  \right|^2} - 1\right].
	\label{e15}
	\end{eqnarray}
	With taking several specific values, we have the mean photon numbers of common states:
	${{\cal N}_{\rm t}} = n_{\rm th}$ for thermal states ($\left| \alpha  \right|=r=0$);
	${{\cal N}_{\rm c}} = {\left| \alpha  \right|^2}$  for coherent states ($n_{\rm th}=r=0$);
	${{\cal N}_{\rm sv}} = {\sinh ^2}r$ for squeezed vacuum states ($\left| \alpha  \right|=n_{\rm th}=0$).
	Moreover, the photon number distribution of a Gaussian state can be calculated as \cite{Yi_min_1997}
	\begin{eqnarray}
	{p_{\rm{g}}}\left( n \right) = \left\langle n \right|{\rho _{\rm{g}}}\left| n \right\rangle  = \frac{2}{{\sqrt {\cal A} }}{e^{ - {\cal B}}}n!{{\cal C}^{2n}}\sum\limits_{k = 0}^n {\frac{{{{\cal C}^{ - 2k}}{{\cal D}^k}}}{{k!{{\left[ {\left( {n - k} \right)!} \right]}^2}}}} H_{n - k}^{}\left( {\frac{\cal E} {\cal C} } \right)H_{n - k}^{}\left( {\frac{{{{\cal E} ^ * }}}{\cal C} } \right)
	\end{eqnarray}
	with
	\begin{align}
	{\cal A} &= \left( {1 + \mu {e^{2r}}} \right)\left( {1 + \mu {e^{ - 2r}}} \right),\\
	{\cal B} &= 2{{\left[ {\left( {\mu  + {e^{ - 2r}}} \right){{\left| \alpha  \right|}^2}{{\cos }^2}\phi  + \left( {\mu  + {e^{2r}}} \right){{\left| \alpha  \right|}^2}{{\sin }^2}\phi } \right]} \mathord{\left/
			{\vphantom {{\left[ {\left( {\mu  + {e^{ - 2r}}} \right){{\left| \alpha  \right|}^2}{{\cos }^2}\phi  + \left( {\mu  + {e^{2r}}} \right){{\left| \alpha  \right|}^2}{{\sin }^2}\phi } \right]} A}} \right.
			\kern-\nulldelimiterspace} A},\\
	{\cal C} &= \sqrt {A\mu \sinh \left( {2r} \right)},\\
	{\cal D} &= {{\left( {{\mu ^2} - 1} \right)} \mathord{\left/
			{\vphantom {{\left( {{\mu ^2} - 1} \right)} A}} \right.
			\kern-\nulldelimiterspace} A},\\
	{\cal E} &= {{\left[ {\left( {{e^{ - r}} + \mu {e^r}} \right)\left| \alpha  \right|\cos \phi  + i\left( {{e^r} + \mu {e^{ - r}}} \right)\left| \alpha  \right|\sin \phi } \right]} \mathord{\left/
			{\vphantom {{\left[ {\left( {{e^{ - r}} + \mu {e^r}} \right)\left| \alpha  \right|\cos \phi  + i\left( {{e^r} + \mu {e^{ - r}}} \right)\left| \alpha  \right|\sin \phi } \right]} A}} \right.
			\kern-\nulldelimiterspace} A},
	\end{align}
	where $\mu  = 2{n_{{\rm{th}}}} + 1$, and ${H_n}\left(  \cdot  \right)$ is the $n$th-degree Hermite polynomial, defined as \cite{abramowitz1965handbook}
	\begin{eqnarray}
	{H_n}\left( x \right) = \sum\limits_j {\frac{{{{\left( { - 1} \right)}^j}n!}}{{j!\left( {n - 2j} \right)!}}} {\left( {2x} \right)^{n - 2j}}.
	\end{eqnarray}
	
	Based on above results, the QFIs of some special Gaussian states will be discussed in the following.
	We first consider thermal states.
	They can be thought of as the cornerstone of Gaussian states, in that each Gaussian state can decompose into thermal states. 
	A thermal state is the superposition of number states whose weight follow Bose-Einstein statistics,
	its photon number distribution is given by
	\begin{eqnarray}
	{p_{\rm{t}}}\left( n \right) = \frac{{n_{{\rm{th}}}^n}}{{{{\left( {{n_{{\rm{th}}}} + 1} \right)}^{n + 1}}}}.
	\end{eqnarray}

	Now we turn our attention to coherent states, a class of important resources in quantum information processing.
	They are the eigenstates of the annihilation operators, and form an overcomplete basis due to their nonorthogonality.
	The probability of finding $n$ photons in a coherent state can be written as
	\begin{eqnarray}
	{p_{\rm c}}\left( n \right) = \left\langle n \right|{\rho _{\rm cs}}\left| n \right\rangle  = \exp \left( { - {{\left| \alpha  \right|}^2}} \right)\frac{{{{\left| \alpha  \right|}^{2n}}}}{{n!}}.
	\end{eqnarray}

	Next, we consider squeezed vacuum states.
	There are only even photons appear in photon number distribution of a squeezed vacuum, since it arises from parametric down-conversion processes.
	The specific number distribution is as follows:
	\begin{eqnarray}
	{p_{\rm sv}}\left( {{\rm{2}}n} \right) = \left\langle {{\rm{2}}n} \right|{\rho _{\rm sv}}\left| {{\rm{2}}n} \right\rangle  = \frac{{\left( {2n} \right)!}}{{{2^{2n}}n!}}\frac{{{{\left( {\tanh r} \right)}^{2n}}}}{{\cosh r}}.
	\end{eqnarray}

\begin{figure}[htbp]
		\centering\includegraphics[width=0.48\textwidth]{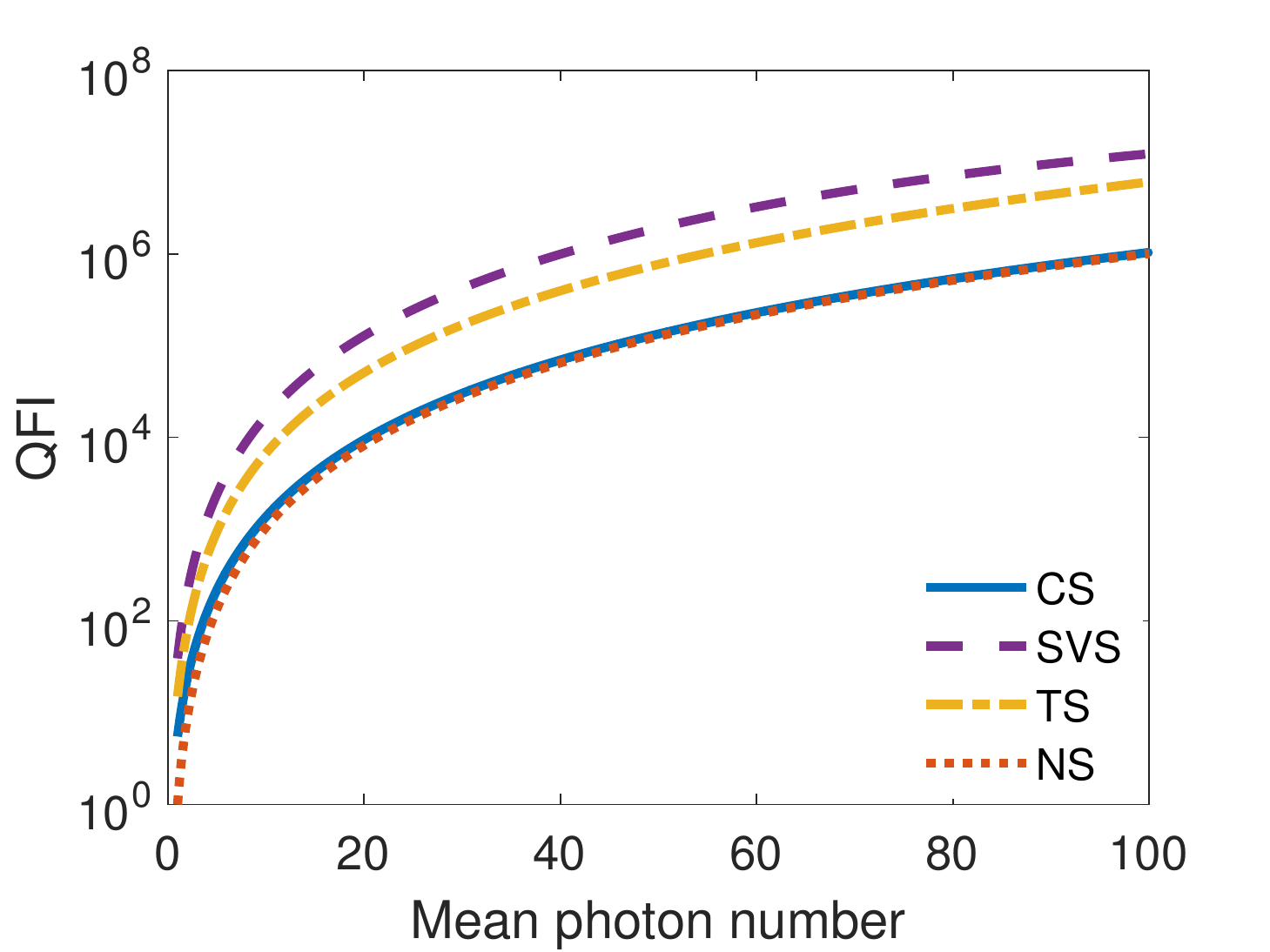}
		\caption{The QFI, ${\cal F}_{\rm Q}$, against mean photon number, ${\cal N}$, with various Gaussian inputs in mode $A$. Figure legends: CS, coherent states; SVS, squeezed vacuum states; TS, thermal states; NS, number states.}
		\label{f2}
	\end{figure}

	So far, we have revisited three states: thermal states, coherent states, and squeezed vacuum states.
	The photon number distributions of above three kinds of states contain merely a single tunable parameter.
	In Fig. \ref{f2}, we give the dependence of their QFIs on the mean photon number.
	As a reference, we also plot the QFI of a number state in terms of Eq. (\ref{e7}).
	One can find that the QFI of the coherent state is approximately equal to that of the number state. 
	The squeezed vacuum or thermal state shows better QFI when compared with the coherent state or number state; meanwhile, the squeezed vacuum is an optimal state among these states.

\begin{figure}[htbp]
		\centering\includegraphics[width=0.48\textwidth]{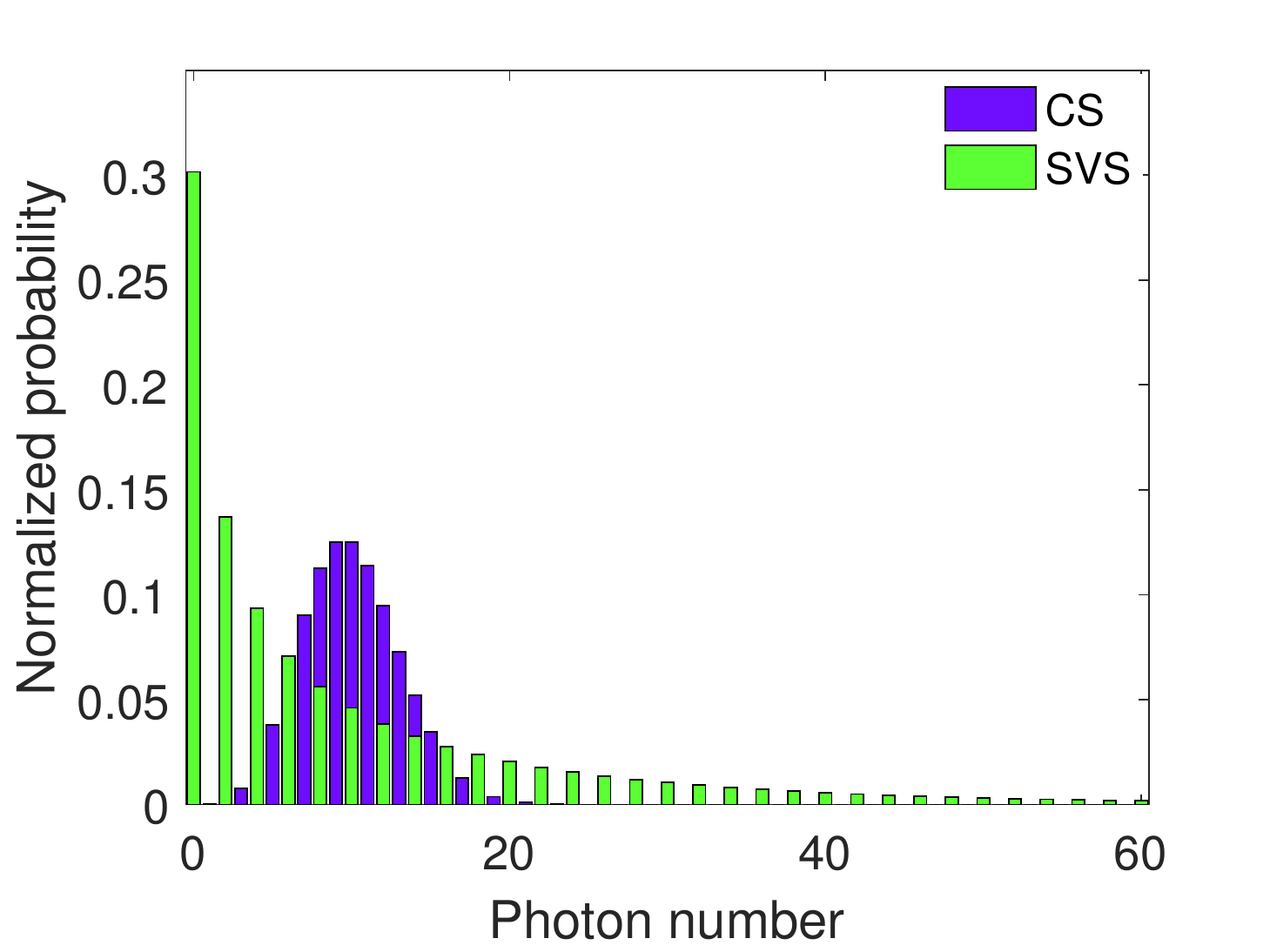}
		\centering\includegraphics[width=0.48\textwidth]{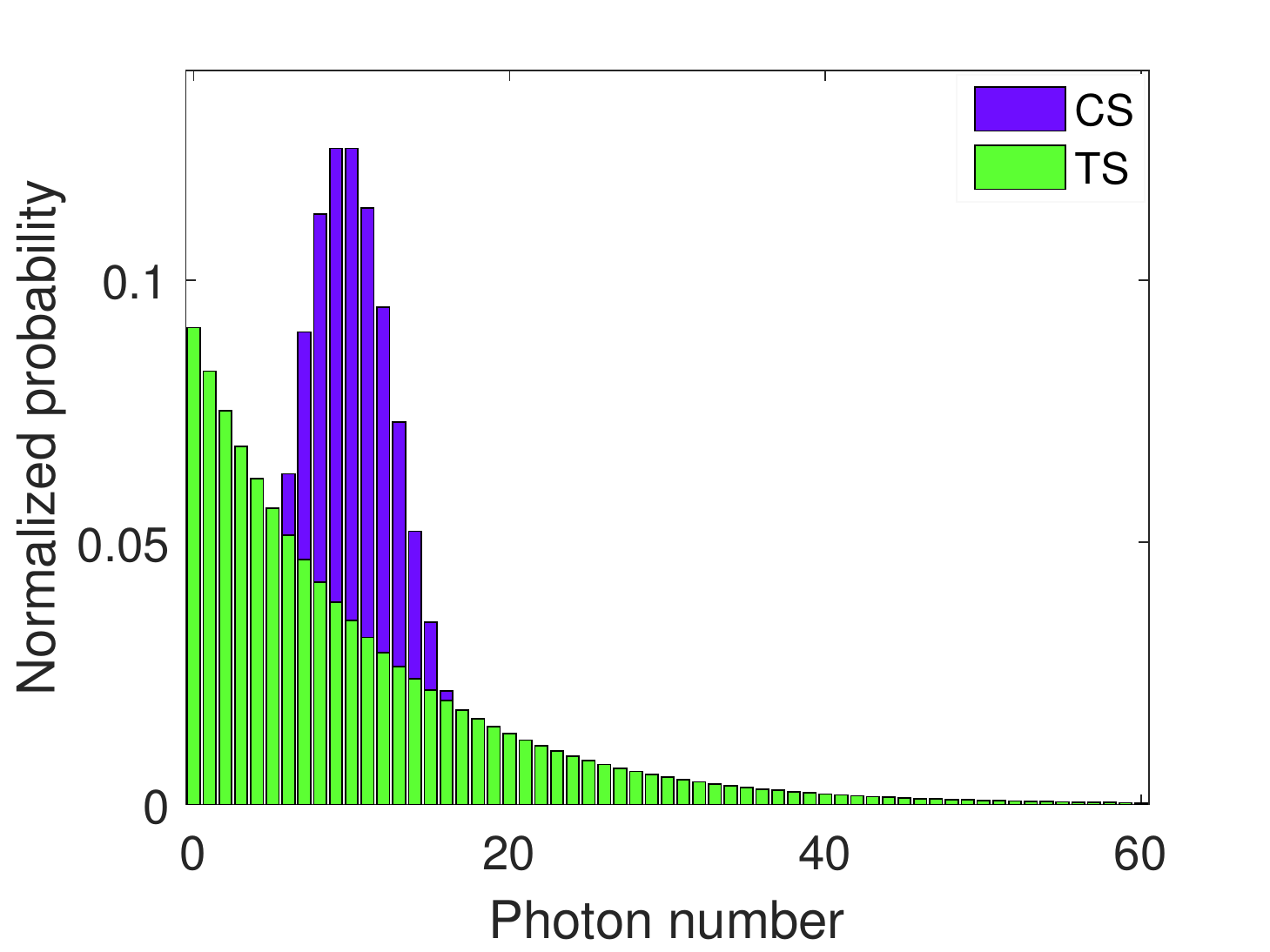}
		\caption{The normalized probability, $p_n$, against photon number, $n$, with ${\cal N}_{\rm sv}={\cal N}_{\rm t}={\cal N}_{\rm c}=10$. Figure legends: CS, coherent states; SVS, squeezed vacuum states; TS, thermal states. (a) Coherent states versus squeezed vacuum states. (b) Coherent states versus thermal states.}
		\label{f3}
	\end{figure}

	To find out the reason for advantages stemming from squeezed vacuum and thermal states, we provide the photon number distributions of these states in Fig. \ref{f3}.
	It can be seen from Fig. \ref{f3}(a) that the photon number distribution of the squeezed vacuum is more dispersed than that of the coherent state.
	As a consequence, the squeezed vacuum has many high-photon probabilities in its photon number distribution.
	Although these probabilities are small, they contribute the most of the QFI.
	This is because the QFI of a number state grows exponentially with the increase of photon number, as shown in Eq. (\ref{e7}).
	In Fig. \ref{f3}(b), the thermal state shows the similar trend with the squeezed vacuum in Fig. \ref{f3}(a).
	That is, there is the same reason why the QFI of the thermal state is better than that of the coherent state.
	Related to this, upon removing zero probabilities of all odd photon numbers from the squeezed vacuum, one can find that it is similar in photon number distribution to the thermal state \cite{doi:10.1080/09500349314551131}.

	Until now, squeezed vacuum states are the optimal input.
	At the last part of this section, we calculate the QFI of a more complex Gaussian state$\--$squeezed coherent state.
	They are once known as two-photon coherent states in history \cite{PhysRevA.13.2226}.
	The probability of emerging $n$ photons in a squeezed coherent state is 
	\begin{align}
	{p_{{\rm{sc}}}}\left( n \right) = \frac{{{{\tanh }^n}r}}{{{2^n}n!\cosh r}}\exp \left[ { - {{\left| \alpha  \right|}^2} - \frac{1}{2}\left( {{\alpha ^{ * 2}} + {\alpha ^2}} \right)\tanh r} \right]{\left| {{H_n}\left( {\frac{{\alpha  + {\alpha ^ * }\tanh r}}{{\sqrt {2\tanh r} }}} \right)} \right|^2}
	\end{align}

	By taking the value of $n_{\rm th}$ in Eq. (\ref{e15}) as 0, one gets the mean photon number of a squeezed coherent state
	\begin{eqnarray}
	{{\cal N}_{\rm sc}} ={\left| \alpha  \right|^2}+\sinh^2{r}
	\label{e27}
	\end{eqnarray}
	It is exactly equal to the sum of the mean photon number of a coherent state and that of a squeezed vacuum.
	Hence, for a given total photon number, we can determinate a squeezed coherent with two parameters: coherent weight,  $z={\cal N}_{\rm c}/{\cal N}_{\rm sc} $; and relative phase, $\phi-\chi$.
	For simplicity, here we select $\chi=0$ and, as a result, $\phi$ stands for the relative phase.

\begin{figure}[htbp]
		\centering\includegraphics[width=0.48\textwidth]{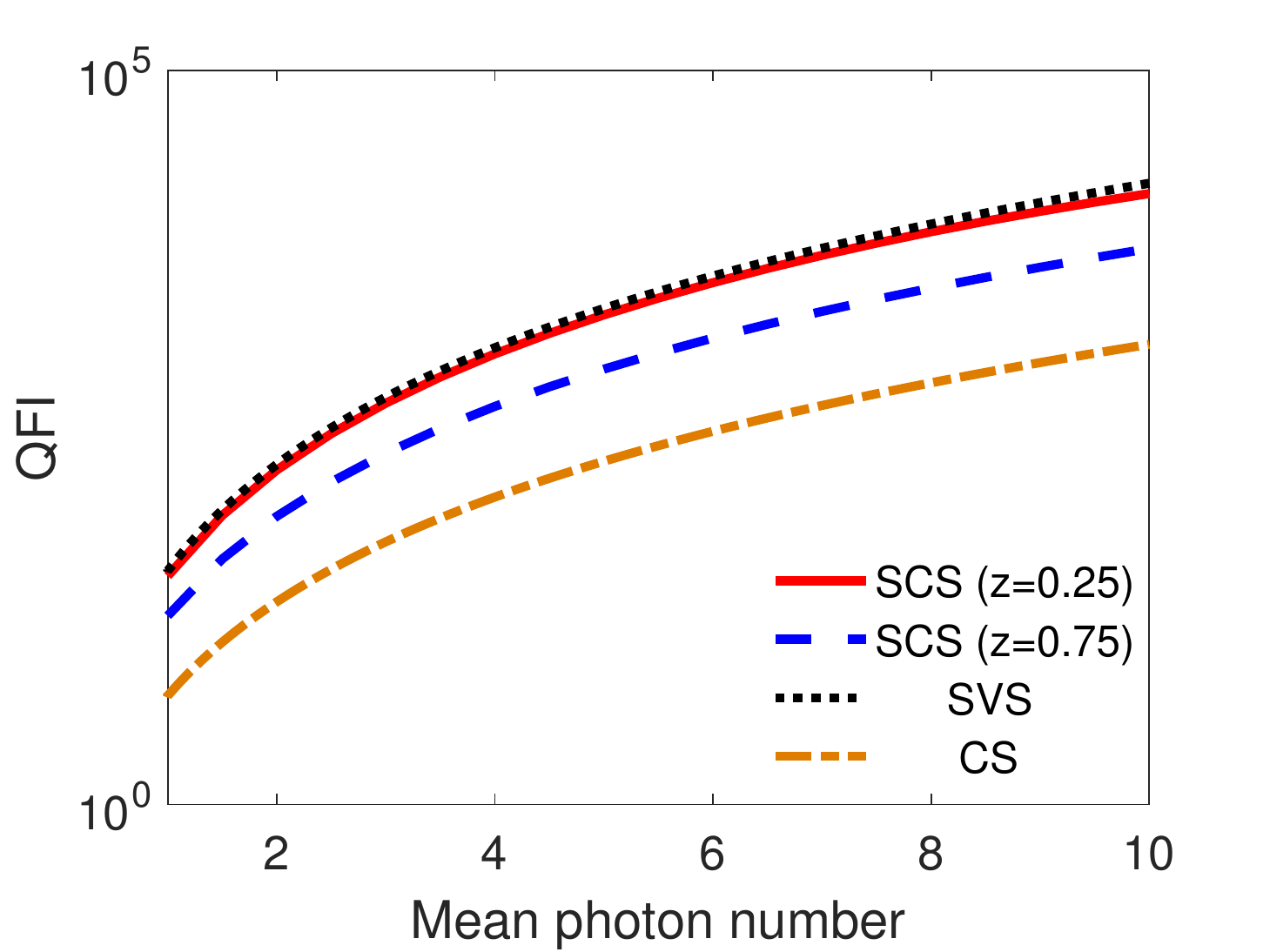}
		\centering\includegraphics[width=0.48\textwidth]{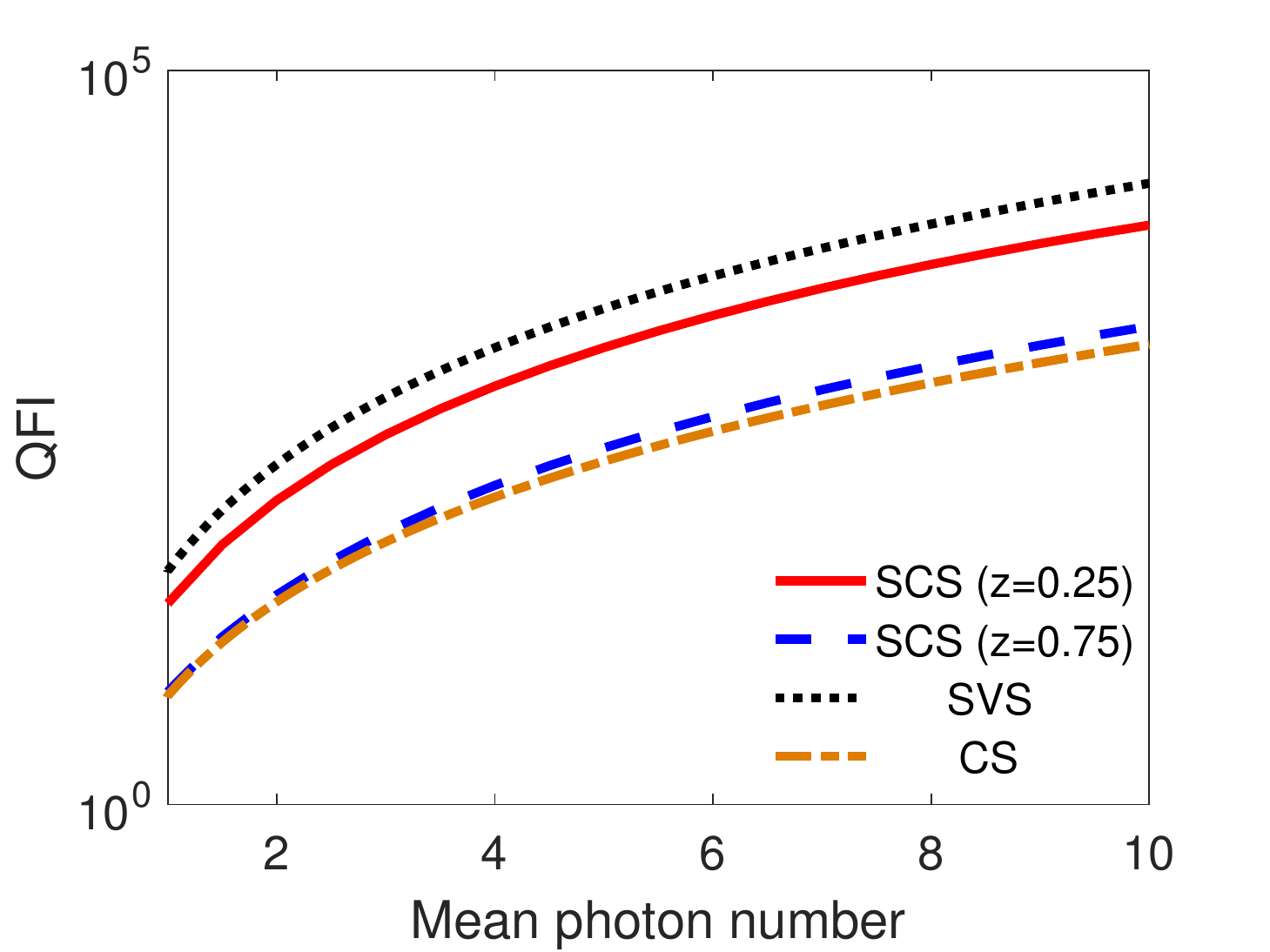}
		\caption{The QFI, ${\cal F}_{\rm Q}$, against mean photon number, ${\cal N}$, with various Gaussian inputs in mode $A$, where $z={\cal N}_{\rm c}/{\cal N}_{\rm sc} $. (a) Squeezed coherent states with $\phi = \pi/2$, super-Poissonian distribution. (b) Squeezed coherent states with $\phi = 0$, sub-Poissonian distribution. Figure legends: SCS, squeezed coherent states; SVS, squeezed vacuum states; CS, coherent states.}
		\label{f4}
	\end{figure}

	In Fig. \ref{f4}, we plot the QFIs of squeezed coherent states with different coherent weights and relative phases.
	The QFIs of the coherent state and squeezed vacuum are also provided as a reference.
	Due to the complexity of Hermite polynomials, we only show these states within 10 photons on average.
	Figure \ref{f4} indicates that the squeezed vacuum remains the optimal input to date.
	With the increase of coherent weight, the QFI of the squeezed coherent state gradually degenerates into that of the coherent state.
	Regarding different relative phases, these squeezed coherent states give the different QFIs.
	By comparison, the super-Poissonian distribution ($\phi=\pi/2$) has sensitivity advantage over the sub-Poissonian distribution ($\phi=0$).
	Overall, the QFI of the squeezed coherent state lies between those of the coherent state and squeezed vacuum.

	\subsection{Squeezed number states}
	In this section, the discussion moves on to a non-Gaussian state$\--$squeezed number state \cite{PhysRevA.40.2494,PhysRevA.45.2044}.
	With regard to a number state $\left| m \right\rangle $, it can be expressed as ${\left| \psi  \right\rangle }=
	\hat S\left( r \right)\left| m \right\rangle $, the corresponding mean photon number is found to be  
	\begin{eqnarray}
	{{\cal N}_{{\rm{sn}}}} = m\cosh \left( {2r} \right) + {\sinh ^2}r.
	\end{eqnarray}
	Unlike Eq. (\ref{e27}), it is not equal to the sum of the mean photon number of a number state and that of a squeezed vacuum.
	
	The photon number distribution of a squeezed number state is given by
	\begin{eqnarray}
	{p_{{\rm{sn}}}}\left( n \right) = \frac{{m!n!}}{{{{\left( {\cosh r} \right)}^{2n + {\rm{1}}}}}}{\left( { \frac{1}{2}\tanh r} \right)^{m - n}}{\cos ^2}\frac{{\left( {m - n} \right)\pi }}{2}{\cal G}\left( {r,m,n} \right)
	\label{e29}
	\end{eqnarray}
	with
	\begin{eqnarray}
	{\cal G}\left( {r,m,n} \right) = {\left| {\sum\limits_j {\frac{{{{\left( { - 1} \right)}^j}}}{{j!\left( {n - 2j} \right)!\left[ {j + {{\left( {m - n} \right)} \mathord{\left/
									{\vphantom {{\left( {m - n} \right)} 2}} \right.
									\kern-\nulldelimiterspace} 2}} \right]!}}{{\left( {\frac{1}{2}\sinh r} \right)}^{2j}}} } \right|^2}.
	\end{eqnarray}
	The cosine term ${\cos ^2}\left[ {{{\left( {m - n} \right)\pi } \mathord{\left/
				{\vphantom {{\left( {m - n} \right)\pi } 2}} \right.
				\kern-\nulldelimiterspace} 2}} \right]$ is not physical reality, but a mathematical means of screening the parity of photon number.
	Further, the parity of photon number distribution in the squeezed number state hinges on the parity of the number state.
	That is, to every odd number $m$ there corresponds a squeezed number state containing only odd photon probabilities, and vice versa.
	
	Here we merely take account of a squeezed one-photon state.
	According to Eq. (\ref{e29}), its photon number distribution can be obtained,
	\begin{eqnarray}
	{p_{{\rm{sn1}}}}\left( {2n + 1} \right) = \frac{{\left( {2n + 1} \right)!}}{{{{\cosh }^3}r{{\left( {n!} \right)}^2}}}{\left( { - \frac{1}{2}\tanh r} \right)^{2n}}.
	\end{eqnarray}

	\begin{figure}[htbp]
		\centering\includegraphics[width=0.48\textwidth]{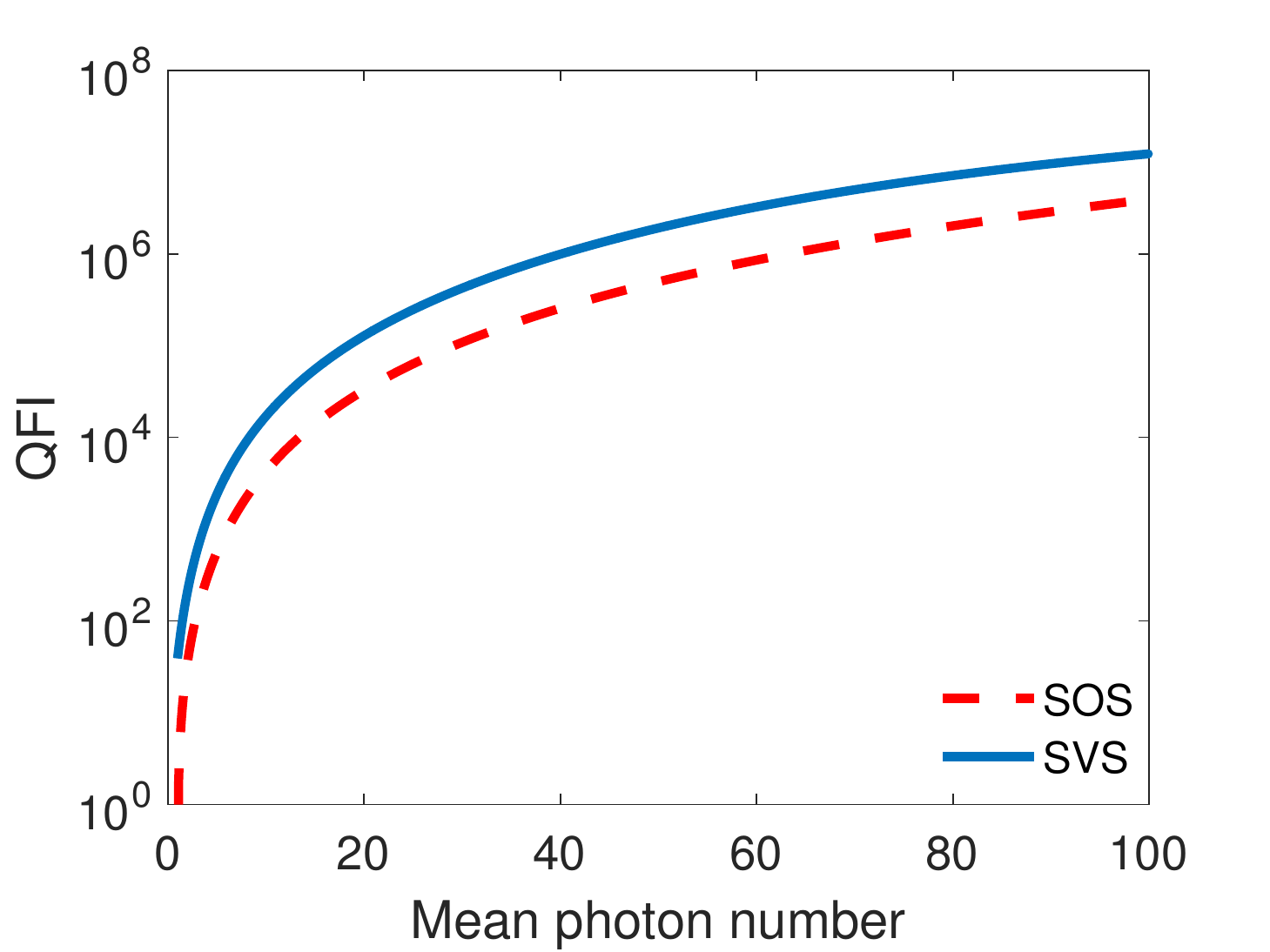}
		\centering\includegraphics[width=0.48\textwidth]{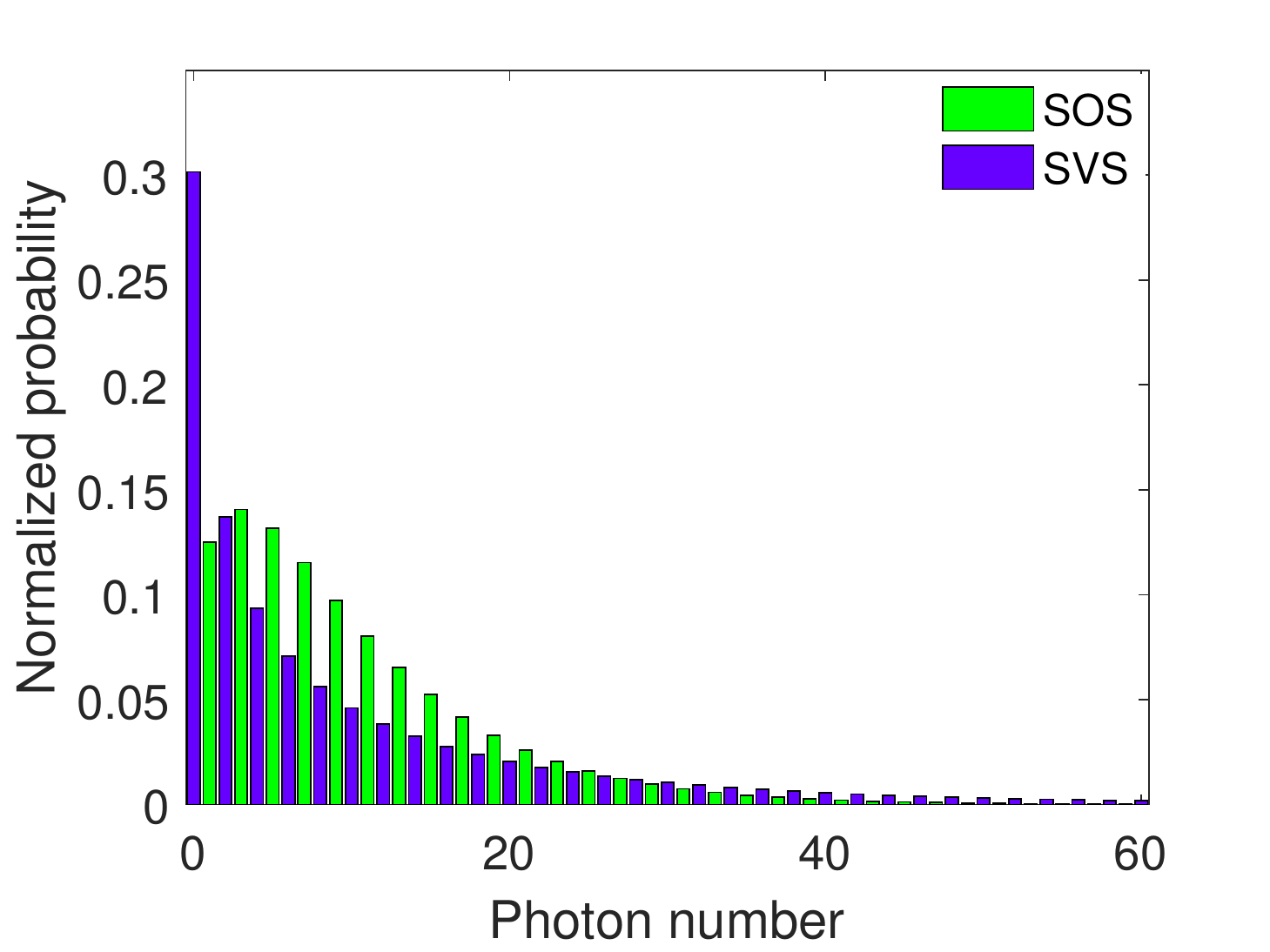}
		\caption{(a) The QFI, ${\cal F}_{\rm Q}$, against mean photon number, ${\cal N}$, with squeezed vacuum and squeezed one-photon inputs in mode $A$. (b) The normalized probability, $p_n$, against photon number, $n$, with ${\cal N}_{\rm sv}={\cal N}_{\rm so}=10$. Figure legends: SVS, squeezed vacuum states; SOS, squeezed one-photon states.}
		\label{f5}
	\end{figure}

	The QFI and photon number distribution of the squeezed one-photon state are shown in Fig. \ref{f5}, as a comparison, we also give those of the squeezed vacuum.
	Figure \ref{f5}(a) suggests that the QFI of the squeezed vacuum is superior to that of the squeezed one-photon state.
	From Fig. \ref{f5}(b), we can find that, as mentioned earlier, the QFI advantage of the squeezed vacuum originates from high-photon probabilities in its photon number distribution.
	In addition, what the results in Figs. \ref{f4} and \ref{f5} reveal is that the QFI of a two-parameter state (squeezed coherent state or squeezed number one) lies between those of two corresponding single-parameter ones.

	\subsection{Schr\"odinger cat states}
	In this section, we report on another important class of non-Gaussian states, Schr\"odinger cat states.
	The form of a general Schr\"odinger cat state \cite{XIA1989281,doi:10.1080/09500349314551131} can be expressed as 
	\begin{equation}
	\left| {{\psi _3}} \right\rangle  = {\cal H}\left( {\left| \alpha  \right\rangle  + {e^{i\delta }}\left| { - \alpha } \right\rangle } \right)
	\end{equation}
	where $\delta$ is a tunable phase, and $\cal H$ is responsible for normalization,
	\begin{equation}
	{\cal H} = \frac{1}{{\sqrt {2\left[ {1 + \exp \left( { - 2{{\left| \alpha  \right|}^2}} \right)\cos \delta } \right]} }}.
	\end{equation}
	In general, there are three kinds of Schr\"odinger cat states in terms of different values of $\delta$: $\delta  = 0$, even coherent states (ECSs); $\delta  = \pi $, odd coherent states (OCSs); $\delta  = {\pi  \mathord{\left/
			{\vphantom {\pi  2}} \right.
			\kern-\nulldelimiterspace} 2}$, Yurke-Stoler coherent states (YSCSs).

	By using the inner product between two coherent states
	\begin{equation}
	\left\langle {\alpha }
	\mathrel{\left | {\vphantom {\alpha  \beta }}
		\right. \kern-\nulldelimiterspace}
	{\beta } \right\rangle  = \exp \left( { - \frac{1}{2}{{\left| \alpha  \right|}^2} - \frac{1}{2}{{\left| \beta  \right|}^2} + {\alpha ^ * }\beta } \right),
	\end{equation}
	we can calculate the mean photon number of a Schr\"odinger cat state
	\begin{equation}
	{{\cal N}_{{\rm{cs}}}} = \frac{{1 - \exp \left( { - 2{{\left| \alpha  \right|}^2}} \right)\cos \delta }}{{1 + \exp \left( { - 2{{\left| \alpha  \right|}^2}} \right)\cos \delta }}{\left| \alpha  \right|^2}
	\end{equation}
	Further, by means of Fock basis,  the photon number distribution of an even coherent state can be obtained
	\begin{equation}
	{p_{{\rm{ec}}}}\left( {2n} \right) =  \left\langle {2n} \right|{\rho _{\rm ec}}\left| {2n} \right\rangle = \frac{1}{{\cosh {{\left| \alpha  \right|}^2}}}\frac{{{{\left| \alpha  \right|}^{4n}}}}{{\left( {2n} \right)!}}.
	\label{e36}
	\end{equation}
	Similarly, one can get the photon number distribution of an odd coherent state, 
	\begin{equation}
	{p_{{\rm{oc}}}}\left( {2n + 1} \right) = \left\langle {2n + 1} \right|{\rho _{\rm oc}}\left| {2n + 1} \right\rangle = \frac{1}{{\sinh {{\left| \alpha  \right|}^2}}}\frac{{{{\left| \alpha  \right|}^{4n + 2}}}}{{\left( {2n + 1} \right)!}},
	\end{equation}
	and that of a Yurke-Stoler coherent state,
	\begin{eqnarray}
	{p_{\rm ys}}\left( n \right) = \left\langle n \right|{\rho _{\rm ys}}\left| n \right\rangle  = \exp \left( { - {{\left| \alpha  \right|}^2}} \right)\frac{{{{\left| \alpha  \right|}^{2n}}}}{{n!}}={p_{\rm c}}\left( n \right).
	\label{e38}
	\end{eqnarray}
	It should be noted that, despite the difference in the ket expressions, a Yurke-Stoler coherent state and a coherent state have the same photon number distributions.
	That is, the QFIs of these two states are identical.

	\begin{figure}[htbp]
		\centering\includegraphics[width=0.48\textwidth]{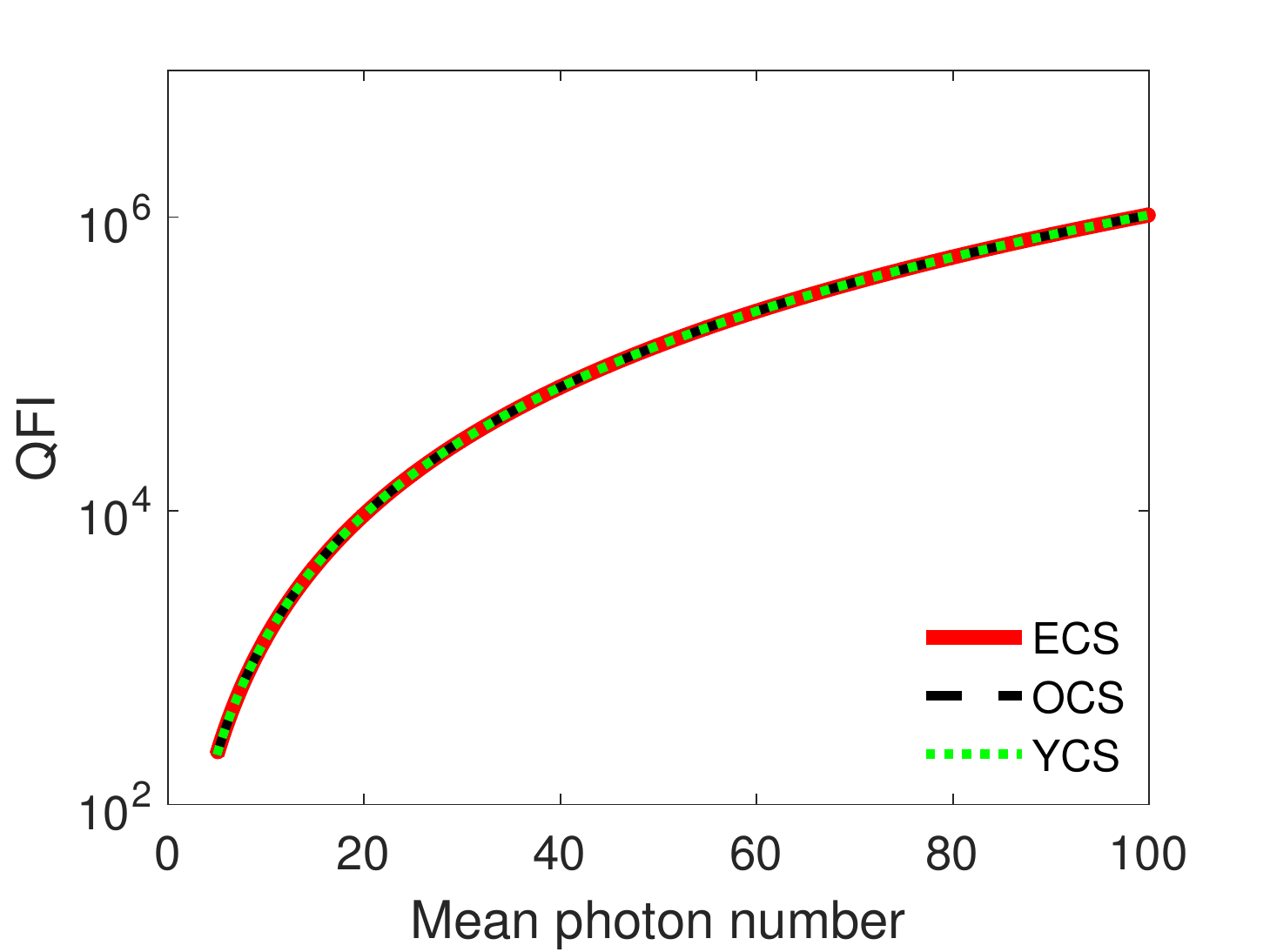}
		\centering\includegraphics[width=0.48\textwidth]{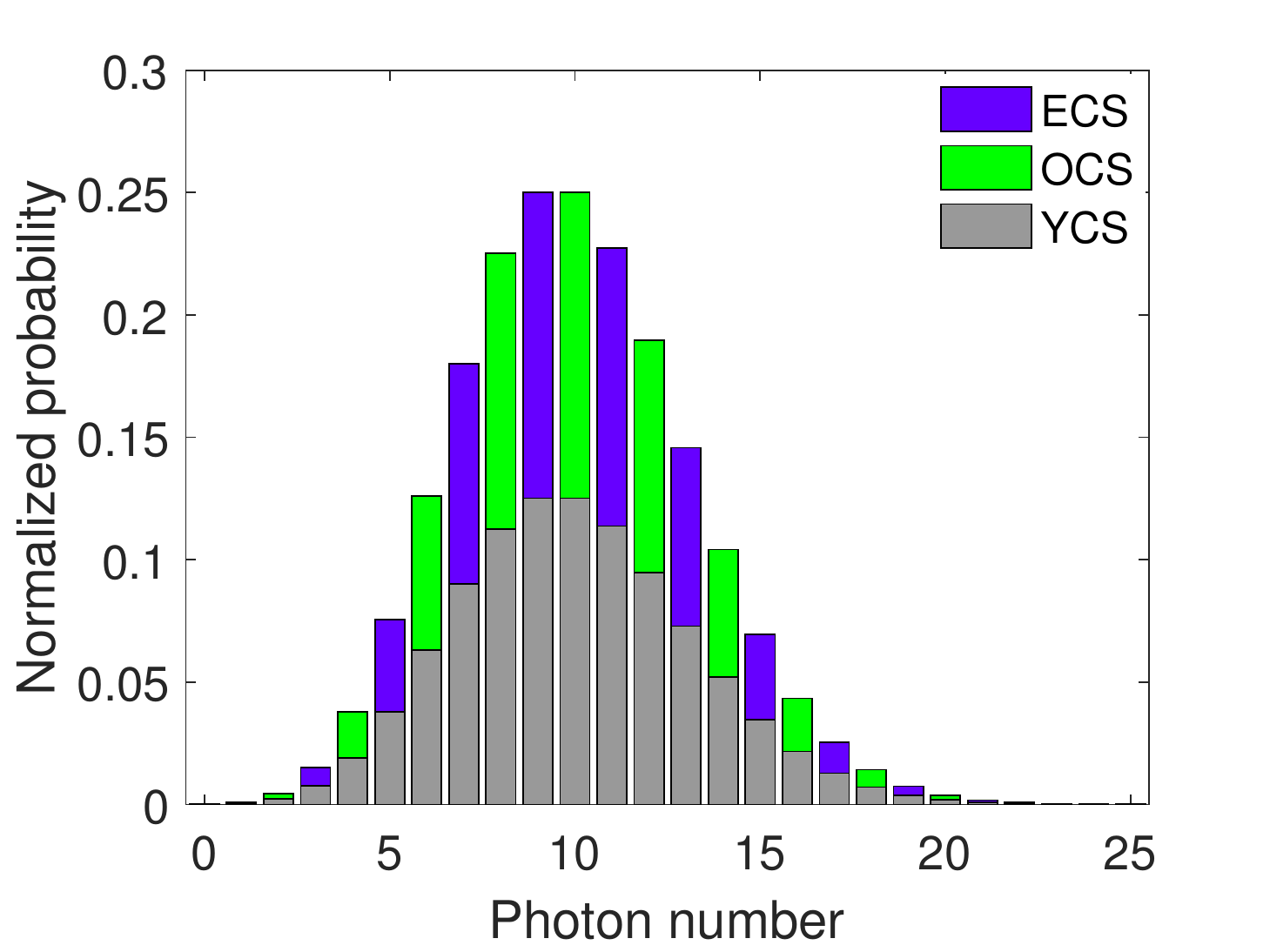}
		\caption{(a) The QFI, ${\cal F}_{\rm Q}$, against mean photon number, ${\cal N}$, with various Schr\"odinger cat inputs in mode $A$. (b) The normalized probability, $p_n$, against photon number, $n$, with ${\cal N}_{\rm ec}={\cal N}_{\rm oc}={\cal N}_{\rm ys}=10$. Figure legends: ECS, even coherent states; OCS, odd coherent states; YCS, Yurke-Stoler coherent states.}
		\label{f6}
	\end{figure}
	
	In Fig. \ref{f6}(a), the QFIs of these cat states are plotted.
	The result manifests that all states approximately exhibit the same QFI.
	To interpret this phenomenon, we give the photon number distributions of them in Fig. \ref{f6}(b).
	As can be seen from the figure, three cat states have the same photon distribution interval.
	Combining Eqs. (\ref{e36}) and (\ref{e38}), we discuss this phenomenon through the analysis of the even coherent state and Yurke-Stoler coherent state.
	Given an even photon number, the probability of the even coherent state is twice as much as that of the Yurke-Stoler coherent state for a large ${\left| \alpha  \right|}^{2}$.
	Meanwhile, there is total lack of odd photon probabilities in the even coherent state.
	For a large ${\left| \alpha  \right|}^{2}$, we have 
	\begin{eqnarray}
	{p_{{\rm{ec}}}}\left( {2n} \right){\cal F}_{\rm Q}\left( {\left| {2n,0} \right\rangle } \right) \approx {p_{{\rm{ys}}}}\left( {2n} \right){\cal F}_{\rm Q}\left( {\left| {2n,0} \right\rangle } \right) +{p_{{\rm{ys}}}}\left( {2n+1} \right){\cal F}_{\rm Q}\left( {\left| {2n+1,0} \right\rangle } \right)
	\end{eqnarray} 
	since ${p_{{\rm{ec}}}}\left( {2n} \right) \approx 2{p_{{\rm{ys}}}}\left( {2n} \right)\approx 2{p_{{\rm{ys}}}}\left( {2n+1} \right)$.
	Further, one can understand why the QFIs of these three cat states are approximately the same. 
	Overall, the squeezed vacuum prevails over these cat states; as a consequence, the squeezed vacuum is the optimal candidate for the input of an interferometer with a nonlinear phase.
	
	\section{Conclusions}
	\label{s4}
	In conclusion, we focus on the conclusive nonlinear phase sensitivity limit for an MZI with a single-mode input.
	Regarding a second-order nonlinear phase inside the interferometer, we discuss the existence or non-existence of shot-noise-style sensitivity limit. 
	The QFI is selected as a benchmark for the estimation of optimality. 
	In order to circumvent possible overestimation induced by the external resources, we take advantage of phase-averaging approach to calculate the QFI of our protocol.
	We consider three kinds of common states, including Gaussian states, squeezed number states, and Schr\"odinger cat states.
	The results indicate that there is no shot-noise-style sensitivity bound on nonlinear phase estimation, and the optimal input is a squeezed vacuum.

	\section*{Acknowledgment}
	This work was supported by the National Natural Science Foundation of China (Grant No. 61701139).

	%\appendix

	\section*{Appendix}
    \label{A}
    In this section, we give the proof of Eq. (\ref{QFI}).
	For a density matrix $\rho  = {\rho _a} \otimes {\rho _b}$, based on the spectral decompositions ${\rho _a} = \sum\nolimits_j {{p_j}\left| {{\psi _j}} \right\rangle } \left\langle {{\psi _j}} \right|$ and ${\rho _b} = \sum\nolimits_m {{q_m}\left| {{\varphi _m}} \right\rangle } \left\langle {{\varphi _m}} \right|$, a well-known expression for the QFI is given by
	\begin{equation}
	{{\cal F}_{\rm Q}} = \sum\limits_k {4{{\cal L}_k}} \left\langle {{\phi _k}} \right|{\hat O^2}\left| {{\phi _k}} \right\rangle  - \sum\limits_{k,l} {\frac{{8{{\cal L}_k}{{\cal L}_l}}}{{{{\cal L}_k} + {{\cal L}_l}}}} {\left| {\left\langle {{\phi _k}} \right|\hat O\left| {{\phi _l}} \right\rangle } \right|^2},
	\label{a1}
	\end{equation}
	where ${{\cal L}_k} = {p_j}{q_m}$, and $\left\{ {\left| {{\phi _k}} \right\rangle  = {{\left| {{\psi _j}} \right\rangle }_A} \otimes {{\left| {{\varphi _m}} \right\rangle }_B}} \right\}$ denotes a complete set basis in the two-mode Hilbert space.

	In our protocol, we have ${\rho _a} = \sum\nolimits_n {{p_n}\left| n \right\rangle } \left\langle n \right|$, ${\rho _b} = \left| 0 \right\rangle \left\langle 0 \right|$, and $\hat O =\hat U_{\rm BS}^\dag {( {\hat a_1^\dag \hat a_1^{}})^2}\hat U_{\rm BS}^{}$.
	Further, the QFI can be expressed as:
	\begin{align}
	\nonumber {{\cal F}_{\rm Q}}\left( {{\bar \rho} } \right) = &\sum\limits_n {4{p_n}} \left\langle {n,0} \right|{\left[ \hat U_{\rm BS}^\dag {( {\hat a_1^\dag \hat a_1^{}} )^2}\hat U_{\rm BS}^{} \right]^2}\left| {n,0} \right\rangle  - \sum\limits_n {4{p_n}} {\left| {\left\langle {n,0} \right|\hat U_{\rm BS}^\dag {( {\hat a_1^\dag \hat a_1^{}} )^2}\hat U_{\rm BS}^{}\left| {n,0} \right\rangle } \right|^2} \\ 
	&- \sum\limits_{n \ne m} {\frac{{8{p_n}{p_m}}}{{{p_n} + {p_m}}}} {\left| {\left\langle {m,0} \right|\hat U_{\rm BS}^\dag {( {\hat a_1^\dag \hat a_1^{}} )^2}\hat U_{\rm BS}^{}\left| {n,0} \right\rangle } \right|^2}. 
	\label{a3}
	\end{align}
	
	One can find that the first two terms can be written as:
	\begin{align}
	\nonumber &\sum\limits_n {4{p_n}} \left\langle {n,0} \right|{\left[ \hat U_{\rm BS}^\dag {( {\hat a_1^\dag \hat a_1^{}})^2}\hat U_{\rm BS}^{} \right]^2}\left| {n,0} \right\rangle  - \sum\limits_n {4{p_n}} {\left| {\left\langle {n,0} \right|\hat U_{\rm BS}^\dag {( {\hat a_1^\dag \hat a_1^{}})^2}\hat U_{\rm BS}^{}\left| {n,0} \right\rangle } \right|^2} \\
	=& \sum\limits_n {{p_n}} {{\cal F}_{\rm Q}}\left( {\left| {{n,0}} \right\rangle } \right).
	\label{a4}
	\end{align}
	The proof of Eq. (\ref{QFI}) goes to prove the following equation:
	\begin{eqnarray}
	\sum\limits_{n \ne m} {\frac{{8{p_n}{p_m}}}{{{p_n} + {p_m}}}} {\left| {\left\langle {m,0} \right|\hat U_{\rm BS}^\dag {{( {\hat a_1^\dag \hat a_1^{}} )}^2}\hat U_{\rm BS}^{}\left| {n,0} \right\rangle } \right|^2} = 0
	\label{proof}
	\end{eqnarray}
	
	To prove this equation, we can use the Heisenberg picture.
	By acting the operator $\hat U_{\rm BS}^{}$ on the input state, we have
	\begin{eqnarray}
	\sum\limits_{n \ne m} {\frac{{8{p_n}{p_m}}}{{{p_n} + {p_m}}}} {\left| {\left\langle {m,0} \right|\hat U_{\rm BS}^\dag {{( {\hat a_1^\dag \hat a_1^{}} )}^2}\hat U_{\rm BS}^{}\left| {n,0} \right\rangle } \right|^2} = \sum\limits_{n \ne m} {\frac{{8{p_n}{p_m}}}{{{p_n} + {p_m}}}} {\left| {\left\langle {{\psi _m}} \right|{{( {\hat a_1^\dag \hat a_1^{}})}^2}\left| {{\psi _n}} \right\rangle } \right|^2}
	\end{eqnarray}
	with
	\begin{eqnarray}
	| \psi_{n} \rangle=\sum_{j=0}^{n} \sqrt{\frac{n !}{j!(n-j)!}}\left(\frac{1}{\sqrt{2}}\right)^{n} | j \rangle | n-j \rangle
	\end{eqnarray}
	The Fock basis $\left| j \right\rangle \left| {n - j} \right\rangle$ of the state $\left| {{\psi _n}} \right\rangle $ is the eigenstate of operator ${\hat a_1^\dag \hat a_1^{}}$; as a result, we get
	\begin{align}
	{( {\hat a_1^\dag \hat a_1^{}})^2}\left| {{\psi _n}} \right\rangle  = \sum\limits_{j = 0}^n  {j^2} \sqrt {\frac{{n!}}{{j!\left( {n - j} \right)!}}} {\left( {\frac{1}{{\sqrt 2 }}} \right)^n} \left| j \right\rangle \left| {n - j} \right\rangle. 
	\end{align}

	Further, the term in Eq. (\ref{proof}) turns out to be
	\begin{align}
	\nonumber \left\langle {{\psi _m}} \right|{( {\hat a_1^\dag \hat a_1^{}})^2}\left| {{\psi _n}} \right\rangle  &= \sum\limits_{j' = 0}^m {{c_{j'}}} \sum\limits_{j = 0}^n {{c_j}} {j^2}\left\langle {{j'}}
	\mathrel{\left | {\vphantom {{j'} j}}
		\right. \kern-\nulldelimiterspace}
	{j} \right\rangle \left\langle {{m - j'}}
	\mathrel{\left | {\vphantom {{m - j'} {n - j}}}
		\right. \kern-\nulldelimiterspace}
	{{n - j}} \right\rangle \\
	&= \sum\limits_{j' = 0}^m {{c_{j'}}} \sum\limits_{j = 0}^n {{c_j}} {j^2}{\delta _{j',j}}{\delta _{m - j',n - j}}.
	\end{align} 
	When $j' = j$ (${\delta _{j',j}} = 1$), it is obvious that $m - j' \ne n - j$ (${\delta _{m - j',n - j}} = 0$) as $m \ne n$.
	In other words, it is not guaranteed that the values of two Kronecker delta are 1 simultaneously.
	Hence, the proof of Eq. (\ref{proof}) is completed.

	Overall, the orthogonality of states $\left\{ {\left| {n,0} \right\rangle } \right\}$ is remained after passing through the first beam splitter because the input of mode $B$ is a vacuum $\left| 0 \right\rangle $.


%	\bibliography{sample}
\end{document}